\documentclass{article}
\usepackage{amssymb,amsmath,graphicx,ucs,hyperref}
\usepackage[utf8x]{inputenc}

\newtheorem{proposition}{Proposition}
\newtheorem{theorem}{Theorem}
\newtheorem{definition}{Definition}
\newtheorem{remark}{Remark}

\newenvironment{proof}{\noindent \emph{Proof. }}{\hfill \hbox{\rlap{$\sqcap$}$\sqcup$}\\}

\title{A Densest ternary circle packing in the plane}

\author{Thomas Fernique}

\date{}

\begin{document}
\maketitle

\begin{abstract}
We consider circle packings in the plane with circles of sizes $1$, $r\simeq 0.834$ and $s\simeq 0.651$.
These sizes are algebraic numbers which allow a compact packing, that is, a packing in which each hole is formed by three mutually tangent circles.
Compact packings are believed to maximize the density when there are possible.
We prove that it is indeed the case for these sizes.
The proof should be generalizable to other sizes which allow compact packings and is a first step towards a general result.
\end{abstract}

\section{Introduction}

A {\em circle packing} in the plane is a set of interior-disjoint circles.
Its {\em density} $\delta$ is defined by
\begin{displaymath}
\delta:=\limsup_{k\to \infty}\frac{\textrm{area in the square $[-k,k]^2$ in the circle interiors}}{\textrm{area of the square $[-k,k]^2$}}.
\end{displaymath}
A central question in circle packings is to find the maximal possible density.
If the circles have all the same size (throughout this paper, by "size" of a circle we mean its radius), it was proven in \cite{FT43} that the density is maximal for the {\em hexagonal compact packings}, where circles are centered on a suitably scaled triangular grid.
For circles of sizes $1$ and $r$, the maximal density has been obtained for seven "magic" values of $r$ \cite{Hep00,Hep03,Ken04}.
These values are specific algebraic numbers which allow a {\em compact packing}, that is, a packing in which each hole is formed by three mutually tangent circles (as in the hexagonal compact packing).
Equivalently, a circle packing is compact if the graph which connects the centers of any two tangent circles is a triangulation.
To the best of our knowledge, no other case is known.
In all these cases, the density is maximized by a compact packing.
Compact packings thus seem to play a crucial role, and we may wonder whether the density is always maximized by a compact packing when the circle sizes allow such a packing\footnote{Clearly, one shall also assume that the densest among the possible compact packing is {\em saturated}, {\em i.e.}, no circle can be further added. Otherwise, adding a circle would yield a more dense non-compact packing.}.

For circles of sizes $1$ and $r$, it was proven in \cite{Ken06} that there are only $9$ values of $r$ which allow a compact packing where circles of both sizes appear: the $7$ above-mentioned "magic" values and two other ones, for which the maximal density is unknown.
For circles of sizes $1$, $r$ and $s$, it was proven in \cite{FHS} that there are $164$ pairs $(r,s)$ which allow a compact packing where circle of the three sizes appear, for all of which the maximal density is unknown.
The case of four or more circle sizes seems to have not yet been considered.

\begin{figure}[hbtp]
\centering
\includegraphics[width=\textwidth]{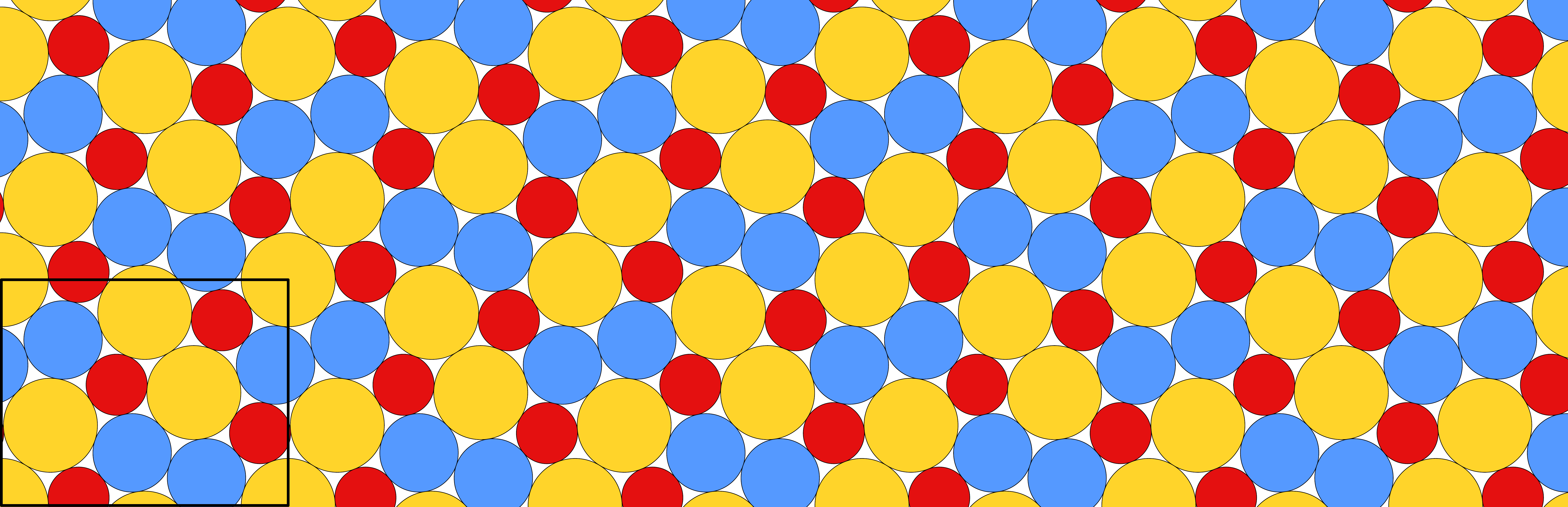}
\caption{
A densest ternary circle packing.
It is compact and periodic (a primitive cell is framed).
}
\label{fig:packing}
\end{figure}

In this paper, we consider one of the $164$ pairs $(r,s)$ which allow a compact packing, namely the roots $r\simeq 0.834$ and $s\simeq 0.651$ of the polynomials
\begin{eqnarray*}
&&4r^8 - 36r^7 - 27r^6 + 162r^5 + 135r^4 - 88r^3 - 73r^2 - 14r + 1,\\
&&89s^8 + 1344s^7 + 4008s^6 - 464s^5 - 2410s^4 + 176s^3 + 296s^2 - 96s + 1.
\end{eqnarray*}
Besides the three homothetic hexagonal compact packings where only one size of circle appear, these values are proven in \cite{FHS} to allow a unique compact packing, depicted in Fig.~\ref{fig:packing} and further referred to as the {\em target packing}.
It has a density $\delta\simeq 0.9093$\footnote{Precisely, one can show that $\delta/\pi\simeq 0.2894$ is root of $129777664x^8 + 11526176768x^7 - 38395680512x^6 + 22192248320x^5 - 41913015856x^4 - 63721188256x^3 + 41344255112x^2 - 80069696280x + 21526627817$.}, what is slightly higher than the density $\pi/\sqrt{12}\simeq 0.9067$ of the hexagonal compact packing.
This is thus the densest compact packing with these circles.
The main result of this paper is that it is also the densest packing at all:

\begin{theorem}
\label{th:main}
The target packing depicted in Fig.~\ref{fig:packing} is the densest packing with these circles.
\end{theorem}

An incidental reason why we chose this packing is that it improves a lower bound given in \cite{FT64} on the largest $q<1$ such that there exists a packing with circles of sizes in $[q,1]$ which is more dense that the hexagonal compact packing.
The bound given in \cite{FT64} is $q\geq 0.6457072159$ whereas the target packing (Fig.~\ref{fig:packing}) yields $q\geq 0.6510501858$.
This bound is increased in \cite{CP} to $q\geq 0.6585340820$ by slightly deforming the target packing.

\section{Strategy}

The strategy to prove Theorem~\ref{th:main} is largely inspired by the one in \cite{Ken04}, with some improvements (different plane decomposition, shorter proofs and more systematic use of the computer).
Given a circle packing, we shall first decompose the plane by a specific triangulation $\mathcal{T}$ of the circle centers (Sec.~\ref{sec:triangulation}).
Then, the {\em excess} $E(T)$ of a triangle $T\in\mathcal{T}$ is defined by
\begin{displaymath}
E(T):=\delta\times\textrm{area}(T)-\textrm{cov}(T),
\end{displaymath}
where $\textrm{area}(T)$ is the area of $T$, $\textrm{cov}(T)$ is the area of $T$ inside the circles centered on the vertices of $T$ and $\delta\simeq 0.9093$ is the density of the target packing (Fig.~\ref{fig:packing}).
Proving that this packing has density at most $\delta$ thus amounts to show
\begin{displaymath}
\sum_{T\in \mathcal{T}}E(T)\geq 0.
\end{displaymath}
For this, we shall define over triangles a {\em potential} $U$ which satisfies two inequalities.
The first one, further referred to as {\em the global inequality}, involves all the triangles of $\mathcal{T}$:
\begin{equation}
\label{eq:global}
\sum_{T\in \mathcal{T}}U(T)\geq 0.
\end{equation}
The second one, further referred to as {\em the local inequality}, involves any triangle $T$ which can appear in $\mathcal{T}$:
\begin{equation}
\label{eq:local}
E(T)\geq U(T).
\end{equation}
The result then trivially follows:
\begin{displaymath}
\sum_{T\in \mathcal{T}}E(T)\geq\sum_{T\in \mathcal{T}}U(T)\geq 0.
\end{displaymath}
Since the global inequality for $U$ is the same as for $E$, it seems we just made things worse by adding a second inequality.
However, we shall choose $U$ so that the global inequality is "not that global", {\em i.e.}, it follows from an inequality on a finite set of finite configurations.
Namely, the potential of a triangle $T$ will be the sum of {\em vertex potentials} $U_v(T)$ defined on each vertex $v\in T$ and {\em edge potentials} $U_e(T)$ defined on each edge $e\in T$ such that, for any vertex $v$ and edge $e$ of any decomposition $\mathcal{T}$:
\begin{equation}
\label{eq:vertex}
\sum_{T\in\mathcal{T}|v\in T} U_v(T)\geq 0
\end{equation}
\begin{equation}
\label{eq:edge}
\sum_{T\in\mathcal{T}|e\in T} U_e(T)\geq 0.
\end{equation}
The first inequality, involving all the triangles sharing a vertex, is proven Section~\ref{sec:global_vertex} (Prop.~\ref{prop:vertex_pot}).
The second one, involving pairs of triangles sharing an edge, is proven Section~\ref{sec:global_edge} (Prop.~\ref{prop:edge_pot}).

The local inequality has then to be proven for each triangle of the decomposition.
We make two cases, depending whether the triangle is near a so-called {\em tight} triangle or not.
The former case, considered Section~\ref{sec:local_tight}, is proven with elementary differential geometry (Prop.~\ref{prop:local_tight}).
The latter case, considered Section~\ref{sec:local}, is proven with a computer by dichotomy (Prop.~\ref{prop:local}).
Theorem~\ref{th:main} shall follow.

\begin{remark}
The proof of Theorem~\ref{th:main} heavily relies on numerical computations.
We always use {\em interval arithmetic}, which allows {\em exact} computations on intervals which contain the quantities of interest (size, density, excess, potential\ldots).
To prove any inequality $A\geq B$ ({\em e.g.} the inequalities~\eqref{eq:local}, \eqref{eq:vertex} and \eqref{eq:edge}), we prove that the left point $\underline{A}$ of the interval containing $A$ is larger than or equal to the right point $\overline{B}$ of the interval containing $B$.
All the computations were performed with the open-source software SageMath \cite{sage} on our modest laptop, an Intel Core i5-7300U with $4$ cores at $2.60$GHz and $15,6$ Go RAM.
\end{remark}

\section{Triangulation}
\label{sec:triangulation}

Given a circle packing, define the {\em cell} of a circle as the set points of the plane which are closer to this circle than to any other.
These cells form a partition of the plane whose dual is a triangulation, referred to as the {\em FM-triangulation} of the packing.
FM-triangulations have been introduced in \cite{FM58} (see also \cite{FT64}) and are also known as {\em additively weighted Delaunay triangulations}.
We shall use them to prove Theorem~\ref{th:main}.
Let us recall a few properties.

\begin{figure}[hbtp]
\centering
\includegraphics[scale=0.35]{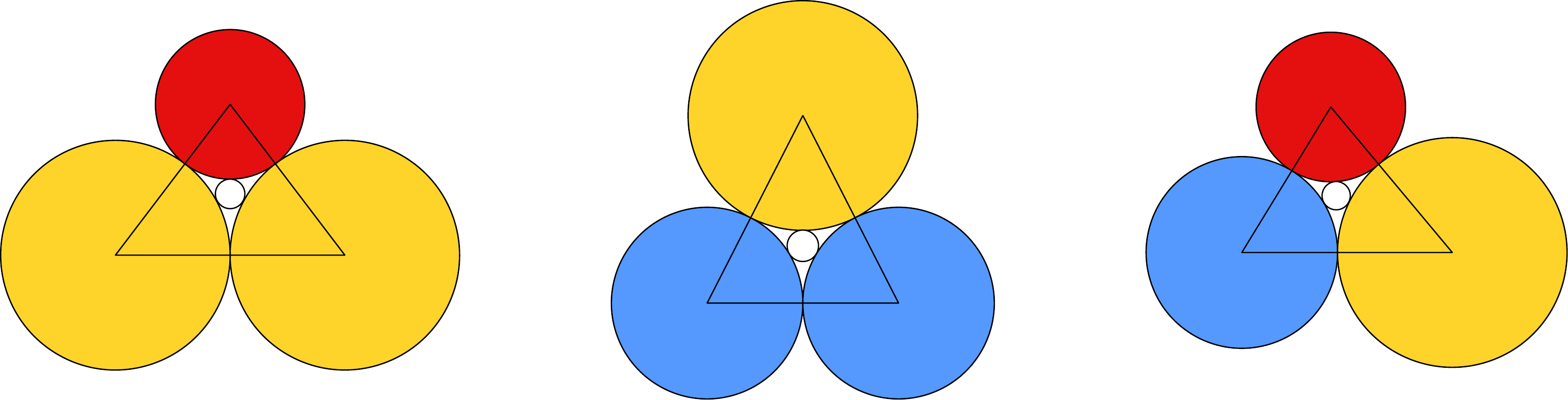}
\caption{The tight triangles appearing in the target depicted (Fig.~\ref{fig:packing}) and their support circle.}
\label{fig:tight}
\end{figure}

If $T$ is a triangle of an FM-triangulation of a circle packing, then one can draw a circle which is interior disjoint from the circles of the packing and tangent to each of the circles of $T$ (see, {\em e.g.}, Fig.~\ref{fig:tight} or \ref{fig:edge_pot}).
It is called {\em support circle} of $T$ and somehow extends the "empty circle property" of the Delaunay triangulation.
In particular, if a packing with circles of sizes at least $s$ is {\em saturated}, {\em i.e.}, no further circle of size $s$ can be added, then any support circle has size strictly less than $s$ and the sector of a circle between two edges of a triangle of its FM-triangulation never crosses the third edge of this triangle.

\begin{figure}[hbtp]
\centering
\includegraphics[scale=0.35]{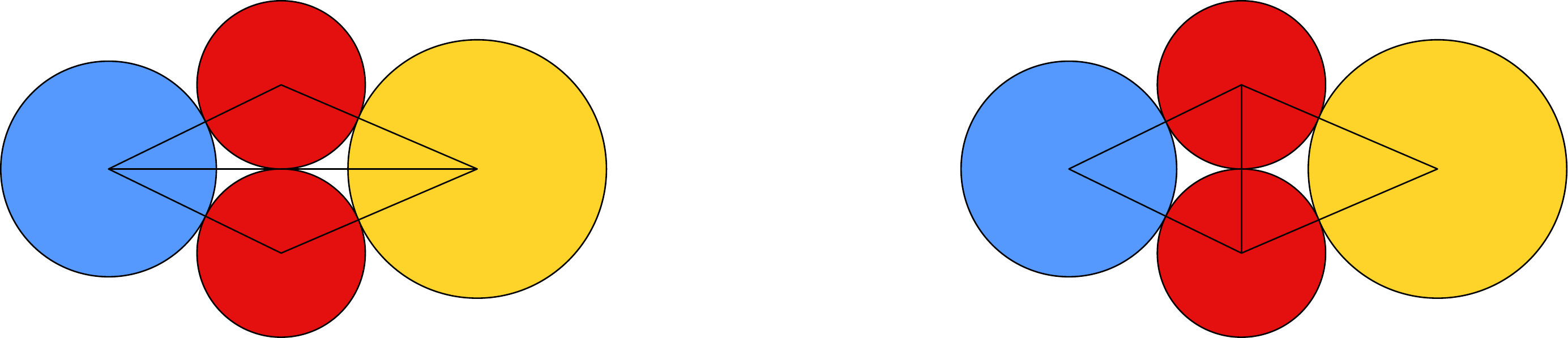}
\caption{
Two adjacent stretched triangles (left) and the two tight triangles obtained by flipping their shared edge (right).
The average density is unchanged.
}
\label{fig:stretched}
\end{figure}

Following \cite{Hep03}, we call {\em tight} a triangle whose circles are mutually adjacent (Fig.~\ref{fig:tight}).
In particular, the FM-triangulation of any compact circle packing contains only tight triangles since there is a support circle in the hole between each three mutually adjacent circles.
Still following \cite{Hep03}, we call {\em stretched} a triangle with one circle tangent to both the two other circles as well as to the line which passes through their centers (Fig.~\ref{fig:stretched}).
Stretched triangles are dangerous because they can be as dense as tight triangles but with a quite different shape.

\section{Global inequality for the vertex potential}
\label{sec:global_vertex}

\begin{figure}[hbtp]
\centering
\includegraphics[width=\textwidth]{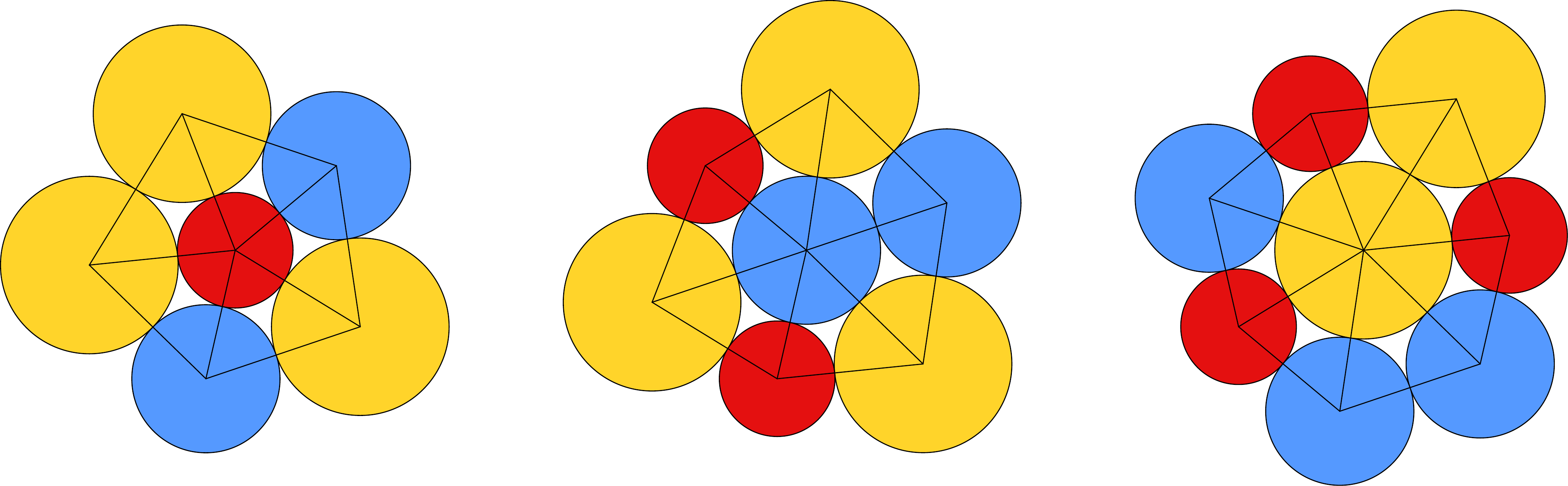}
\caption{
The inequality~\eqref{eq:vertex} enforce $V_{1s1}+4V_{1sr}=0$ around a small circle, $2V_{1rr}+4V_{1rs}=0$ around a medium circle and $V_{r1r}+2V_{1rs}+4V_{r1s}=0$ around a large circle (from left to right).
}
\label{fig:corona_eq}
\end{figure}

We shall first define the vertex potential in tight triangles.
It will depend only on the size $a$, $b$ and $c$ of the circles of the considered triangle.
We denote by $V_{abc}$ (or $V_{cba}$) the potential in the center of the circle of size $b$.
There are thus $18$ quantities to be defined:
\[
\resizebox{\textwidth}{!}{$V_{111}, V_{rrr}, V_{sss}, V_{s1s}, V_{1ss}, V_{r1r}, V_{1rr}, V_{srs}, V_{rss}, V_{1s1}, V_{11s}, V_{1r1}, V_{11r}, V_{rsr}, V_{rrs}, V_{1sr}, V_{r1s}, V_{1rs}.$}   
\]
Consider an FM-triangulation $\mathcal{T}$ of the target packing (Fig.~\ref{fig:packing}).
By summing over the triangles of $\mathcal{T}$, we get
\[
\sum\underbrace{E(T)-U(T)}_{\textrm{$\geq 0$ by \eqref{eq:local}}}=\underbrace{\sum E(T)}_{\textrm{$=0$ by definition of $E$}}-\underbrace{\sum U(T)}_{\textrm{$\geq 0$ by \eqref{eq:global}}}.
\]
This enforces $U(T)=E(T)$ for each of the three tight triangles which appear in $\mathcal{T}$ (Fig.~\ref{fig:tight}).
This yields three equations on the $V_{abc}$'s.
The inequality~\eqref{eq:vertex} yields three further equations on the $V_{abc}$'s, one for each circle size (Fig.~\ref{fig:corona_eq}).
These equations are however not independent because the sum of the excesses of the triangles of $\mathcal{T}$ is zero, hence the sum of the vertex potentials of the tight triangles of a primitive cell of $\mathcal{T}$ is zero\footnote{Namely, this yields $4(V_{1sr}+V_{r1s}+V_{1rs})+(2V_{11s}+V_{1s1})+(2V_{1rr}+V_{r1r})=0$.}.
The triangulation $\mathcal{T}$ thus yields $5$ independent equations on $18$ variables.

We set $U(T)=E(T)$ not only for the three tight triangles which appear in $\mathcal{T}$, but for all the tight triangles: this yields $7$ new equations.
Further, we set $V_{aba}=0$ for any $a\neq b$ in $\{1,r,s\}$: this yields $6$ new equations.
Both settings are arbitrary: they allow to prove Theorem~\ref{th:main} but may need to be revised to adapt the proof to other cases (see discussion Sec.~\ref{sec:further}).
With the $5$ previous equations, it turns out to completely determine the $V_{abc}$'s\footnote{The computed values are ugly and have no particular interest: we do not give them here.}.

We are now in a position to define the vertex potential in any triangle.
The idea is to modify the potential of a tight triangle depending on how much the triangle itself is deformed:

\begin{definition}
\label{def:vertex_pot}
Let $v$ be a vertex of a triangle $T$.
Let $q$ be the size of the circle of center $v$ and $x$ and $y$ the sizes of the two other circles of $T$.
Let $\check T$ be the tight triangle obtained by contracting the edges of $T$ until the three circles become mutually tangent.
The {\em vertex potential} $U_v(T)$ of $v$ is defined by
\[
U_v(T):=V_{xqy}+m_q|\widehat{v}(T)-\widehat{v}(\check T)|,
\]
where $m_q\geq 0$ depends only on $q$, and $\widehat{v}(T)$ and $\widehat{v}(\check T)$ denote the angle in $v$ in $T$ and $\check T$.
\end{definition}

In particular, $U_v(\check T)=V_{xqy}$.
The constant $m_q$ controls the "deviation" in term of the angle changes between $T$ and $\check T$.
The point is to fix it so that the inequality~\eqref{eq:vertex} holds:

\begin{proposition}
\label{prop:vertex_pot}
If $v$ is a vertex of an FM-triangulation of a saturated packing by circles of sizes $1$, $r$ and $s$, then the sum of the vertex potentials of the triangles containing $v$ is nonnegative provided that
\begin{eqnarray*}
m_1&\geq& 0.115610891330759,\\
m_r&\geq& 0.023471932071104,\\
m_s&\geq& 0.022750796636041.
\end{eqnarray*}
\end{proposition}

\begin{proof}
Let $v$ be a vertex of an FM-triangulation $\mathcal{T}$ of a saturated packing by circles of size $1$, $r$ and $s$.
Let $T_1,\ldots,T_k$ be the triangles of $\mathcal{T}$ which contain $v$, ordered clockwise around $v$.
We have:
\begin{eqnarray*}
\sum_i U_v(T_i)
&=&\sum_i U(\check T_i)+m_q\sum_i|\widehat{v}(T_i)-\widehat{v}(\check T_i)|\\
&\geq& \sum_i U(\check T_i)+m_q\left|\sum_i\widehat{v}(T_i)-\sum_i\widehat{v}(\check T_i)\right|.
\end{eqnarray*}
Since the $T_i$'s surround $v$, $\sum_i\widehat{v}(T_i)=2\pi$.
If the coefficient of $m_q$ is nonzero, then the inequality~\eqref{eq:vertex} is thus satisfied in $v$ as soon as
\[
m_q\geq -\frac{\sum_i U(\check T_i)}{\left|2\pi-\sum_i\widehat{v}(\check T_i)\right|}.
\]
This lower bound depends only on the sizes and order of the circles centered on the neighbors of $v$.
In particular, it takes only finitely many values for a fixed $k$.

We claim that $k$ is uniformly bounded thanks to properties of FM-tri\-an\-gu\-la\-tions.
Consider indeed a triangle $T$.
On the one hand, any edge of $T$ has length at most $2+2s$ since one can connect its endpoints to the center of the support circle (of size less than $s$) by two line segments of length at most $1+s$.
On the other hand, the altitude of any vertex of $T$ is at least $s$ since no circle crosses the line which passes through the opposite vertices.
This leads to bound from below by $\tfrac{s}{2+2s}$ the sine of any angle of $T$.
The number $k$ of triangles sharing $v$ is then bounded from above by $\lfloor 2\pi/\arcsin\tfrac{s}{2+2s}\rfloor=31$.

We can thus perform an exhaustive search on a computer to find a lower bound which holds for any $v$.
The claimed values are approximations by excess of the computed values\footnote{We can actually reduce further the number of cases to consider by bounding from below the angle of a triangle depending on the circles of this triangles. It is however only useful to speed up the search, because the cases that give the lower bound on $m_q$ correspond to rather small values of $k$.}.

To finish, we also have to consider the case where $m_q$ has a zero coefficient, that is, when the sum of the angles $\widehat{v}(\check T_i)$ is equal to $2\pi$.
One checks by computing all the angles of all the tight triangles that it happens in only two cases: either the $T_i$'s form one of the three configurations depicted in Fig.~\ref{fig:corona_eq}, or they are $6$ identical equilateral triangles.
In the former case, one has 
$$
\sum_i U_v(T_i)=\sum_i V(\check T_i)=0
$$
because we have chosen the $V_{abc}$ for this purpose (Fig.~\ref{fig:corona_eq}).
In the latter case, one computes
$$
\sum_i U_v(T_i)=6V_{qqq}=2(\delta\sqrt{3}-\tfrac{\pi}{2})q^2>0.
$$
The inequality~\eqref{eq:vertex} is thus satisfied in both cases.
\end{proof}

\section{Global inequality for the edge potential}
\label{sec:global_edge}

\begin{figure}[hbtp]
\centering
\includegraphics[width=\textwidth]{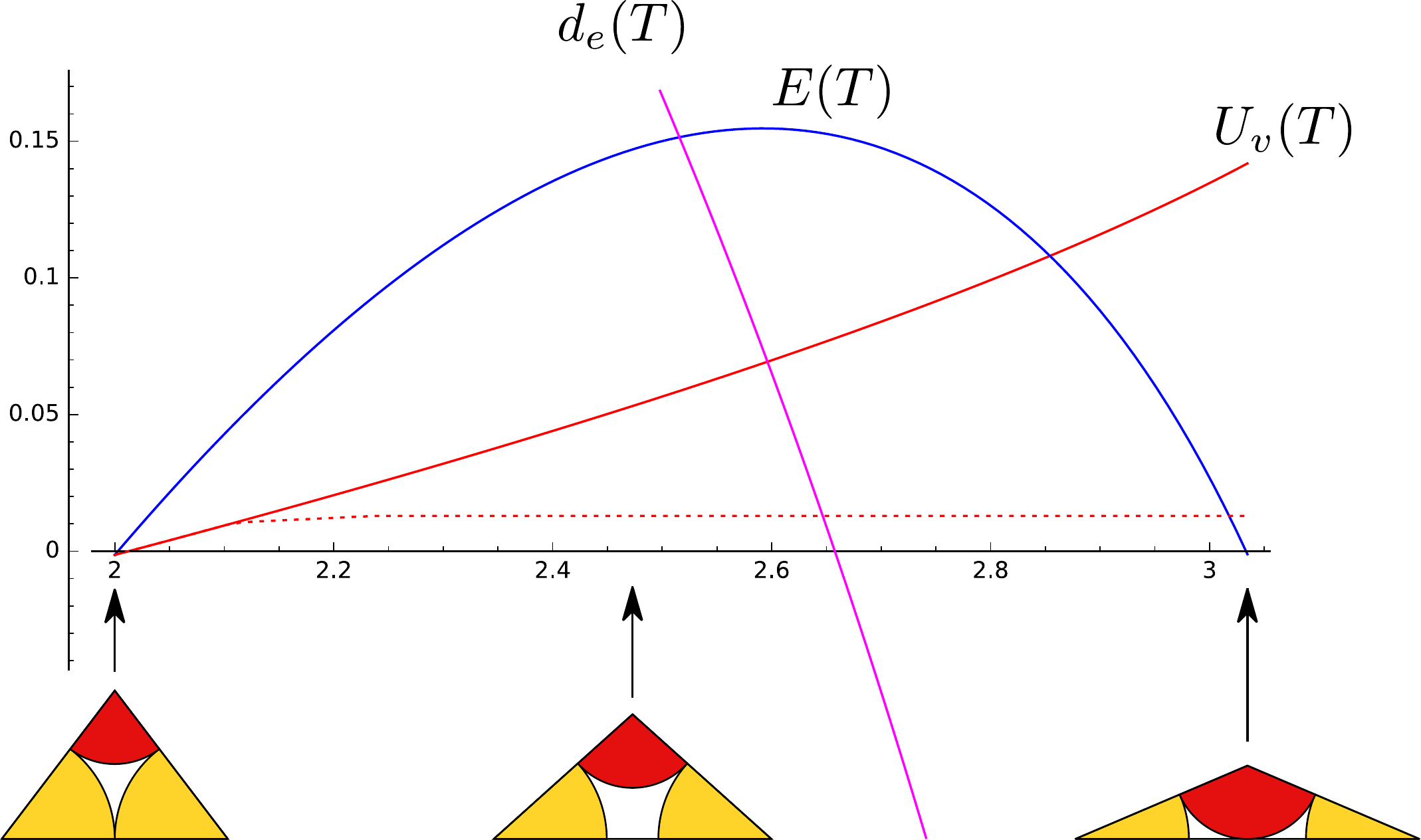}
\caption{
Starting from a tight triangle with two large circle and a small one (bottom left), the edge $e$  between the two large circles is elongated until we get a stretched triangle (bottom right).
The corresponding variations of the excess $E$, the vertex potential $U_v$ and the signed distance $d_e(T)$ are depicted (top).
Near the stretched triangle, the local inequality $E(T)\geq U_v(T)$ fails.
The dashed line is an improvement of $U_v(T)$ discussed in Section~\ref{sec:further}.
}
\label{fig:plot}
\end{figure}

A few randomized trials suggest that the vertex potential satisfy the local inequality~\eqref{eq:local} for triangles which are not too far from tight triangles.
It however fails near stretched triangles, because the excess can become quite small.
The typical situation is depicted in Fig.~\ref{fig:plot}.
The edge potential aims to fix this problem.
The idea is that when a triangle $T$ becomes stretched, its support circle overlaps an adjacent triangle $T'$, imposing a void in $T'$ which increases $E(T')$ and may counterbalance the decrease of $E(T)$.
We shall come back to this in Section~\ref{sec:local}.
Here, we define the edge potential and prove that it satisfies Inequality~\eqref{eq:edge}.

\begin{definition}
\label{def:edge_pot}
Let $e$ be an edge of a triangle $T$.
Let $x$ and $y$ be the sizes of the circles centered on the endpoints of $e$. 
Denote by $d_e(T)$ the signed distance of the center $X$ of the support circle of $T$ to the edge $e$, which is positive if $T$ and $X$ are both on the same size of $e$, or negative otherwise.
The {\em edge potential} $U_e(T)$ of $e$ is defined by
\[
U_e(T):=\left\{\begin{array}{l}
\textrm{$0$ if the edge $e$ is shorter than $l_{xy}$,}\\
\textrm{$q_{xy}\times d_e(T)$ otherwise,}
\end{array}\right.
\]
where $l_{xy}\geq 0$ and $q_{xy}\geq 0$ depend only on $x$ and $y$.
\end{definition}

The constant $l_e$ is the threshold at which $d_e$ has an effect and the coefficient $q_e$ controls the intensity of this effect.
In contrast to the role of $m_q$ for the vertex potential to satisfy Inequality~\ref{eq:vertex} (Prop.~\ref{prop:vertex_pot}), the values of $l_e$ and $q_e$ (and even the circle sizes) do not play a role for the edge potential to satisfy Inequality~\eqref{eq:edge}:

\begin{proposition}
\label{prop:edge_pot}
If $e$ is an edge of an FM-triangulation of a circle packing, then the sum of the edge potentials of the two triangles containing $e$ is nonnegative.
\end{proposition}

\begin{figure}[hbtp]
\centering
\includegraphics[width=0.8\textwidth]{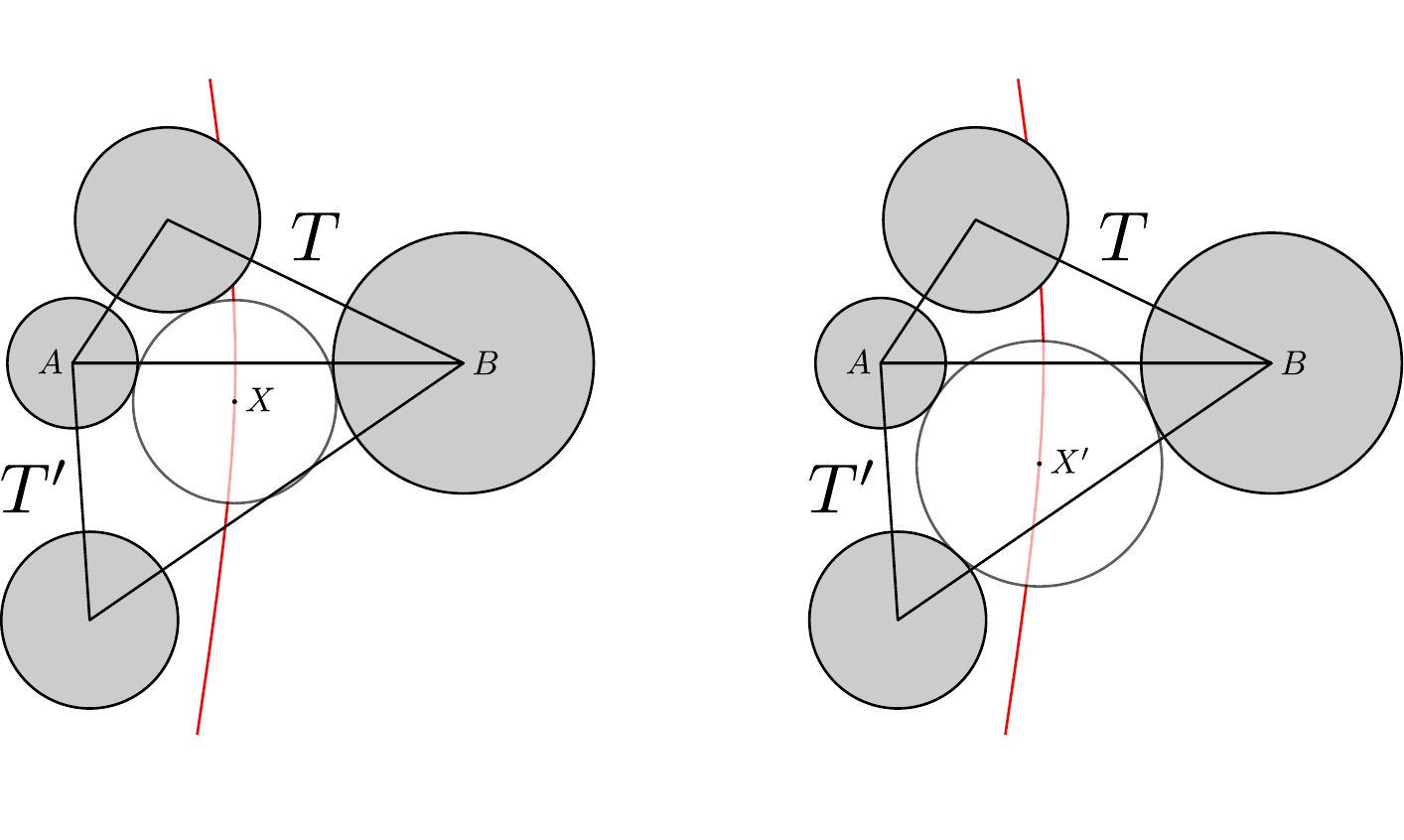}
\caption{Comparatives positions of the centers of the support circles of two adjacent triangles.}
\label{fig:edge_pot}
\end{figure}

\begin{proof}
Consider an edge $e$ shared by two triangles $T$ and $T'$ of an FM-triangulation.
We claim that $d_e(T)+d_e(T')\geq 0$.
If each triangle and the center of its support circle are on the same side of $e$, then it holds because both $d_e(T)$ and $d_e(T')$ are nonnegative.
Assume $d_e(T)\leq 0$, {\em i.e}, $T$ and the center of its support circle are on either side of $e$.
Denote by $A$ and $B$ the endpoints of $e$ and by $a$ and $b$ the sizes of the circles of center $A$ and $B$ (Fig.~\ref{fig:edge_pot}).
The centers of the circles tangent to both circles of center $A$ and $B$ and radii $a$ and $b$ are the points $M$ such that $\overline{AM}-a=\overline{BM}-b$, {\em i.e.}, a branch of a hyperbola of foci $A$ and $B$.
This includes the centers $X$ and $X'$ of the support circles of $T$ and $T'$.
In order to be tangent to the third circle of $T'$, the support circle of $T'$ must have a center $X'$ farther than $X$ from the focal axis.
Since the distances of $X'$ and $X$ to this axis are $-d_e(T)$ and $d_e(T')$, this indeed yields $d_e(T)+d_e(T')\geq 0$.
This proves $U_e(T)+U_e(T')\geq 0$ if $e$ has length at least $l_{ab}$.
If $e$ is shorter, both $U_e(T)$ and $U_e(T')$ are zero and their sum is thus nonnegative.
\end{proof}

\section{Local inequality for $\varepsilon$-tight triangles}
\label{sec:local_tight}

We prove the local inequality~\eqref{eq:local} in a neighborhood of tight triangles.
A triangle is said to be $\varepsilon$-tight if its circles are pairwise at distance at most $\varepsilon$.
Let $\check T$ be a tight triangle with edge length $x_1$, $x_2$ and $x_3$ and denote by $\mathcal{T}_\varepsilon$ the set of $\varepsilon$-triangles with the same circles sizes as $\check T$.
On the one hand, the variation $\Delta E$ of the excess $E$ between $\check T$ and any triangle in $\mathcal{T}_\varepsilon$ satisfies
\[
\Delta E\geq \sum_{1\leq i\leq 3} \min_{\mathcal{T}_\varepsilon}\frac{\partial E}{\partial x_i}\Delta x_i.
\]
On the other hand, assuming that $\varepsilon$ is smaller than the smallest threshold $l_{xy}$ below which the edge potential is zero (so that the potential $U$ is simply the vertex potential, see Def.~\ref{def:vertex_pot}), the variation $\Delta U$ of the potential $U$ between $\check T$ and any triangle in $\mathcal{T}_\varepsilon$ satisfies
\[
\Delta U\leq\sum_{1\leq i\leq 3} \max_{\mathcal{T}_\varepsilon}\frac{\partial U}{\partial x_i}\Delta x_i.
\]
The local inequality $E(T)\geq U(T)$ thus holds over $\mathcal{T}_\varepsilon$ for any $\varepsilon$ such that
\[
\min_{\mathcal{T}_\varepsilon}\frac{\partial E}{\partial x_i}\geq\max_{\mathcal{T}_\varepsilon}\frac{\partial U}{\partial x_i}.
\]
We computed the formulas of the derivatives of $E$ and $U$ with SageMath\footnote{It can be easily do by hand since it mainly amounts to use the cosine theorem to express the angle of a triangle as a function of its edge length, but we are not particularly interested in the formulas.}.
We then use interval arithmetic, once again, to compute the extremal values over $\mathcal{T}_\varepsilon$: each variable $x_i$ is replaced by the interval\footnote{Its endpoints are, as any other numerical quantity, themselves computed with interval arithmetic.} $[r_j+r_k, r_j+r_k+\varepsilon]$, where $r_j$ and $r_k$ denote the sizes of the circles centered on the endpoints of the edge of length $x_i$.
A computation yields:

\begin{proposition}
\label{prop:local_tight}
Fix $m_1=0.12$ and $m_r=m_s=0.03$, which satisfy the lower bound of Prop.~\ref{prop:vertex_pot}.
Then, the local inequality $E(T)\geq U(T)$ holds for any $\varepsilon$-tight triangle of an FM-triangulation of a saturated packing by circles of size $1$, $r$ and $s$ provided that
\[
\varepsilon\leq 0.0561906177650666.
\]
\end{proposition}

\section{Local inequality for all the triangles}
\label{sec:local}

We explicitly define an edge potential such that the local inequality~\eqref{eq:local} holds for any triangle.
Since the global inequality~\eqref{eq:global} result from Prop.~\ref{prop:vertex_pot} and \ref{prop:edge_pot}, this shall prove Theorem~\ref{th:main}.

\begin{proposition}
\label{prop:local}
Fix $m_1=0.12$ and $m_r=m_s=0.03$, which satisfy the lower bound of Prop.~\ref{prop:vertex_pot}.
Fix $\varepsilon=0.056$, which satisfies the corresponding upper bound of Prop.~\ref{prop:local_tight}.
Fix the following values in the definition of the edge potential (Def.~\ref{def:edge_pot}):
\begin{center}
\begin{tabular}{cccccc}
$l_{11}=2.78$, & $l_{1r}=2.60$, & $l_{1s}=2.32$, & $l_{rr}=2.42$, & $l_{rs}=2.14$, & $l_{ss}=1.94$,\\
$q_{11}=0.39$, & $q_{1r}=0.31$, & $q_{1s}=0.30$, & $q_{rr}=0.26$, & $q_{rs}=0.26$, & $q_{ss}=0.25$.
\end{tabular}
\end{center}
Then, the local inequality $E(T)\geq U(T)$ holds for any triangle of an FM-triangulation of a saturated packing by circles of size $1$, $r$ and $s$.
\end{proposition}

\begin{proof}
We shall check the inequality over all the possible triangles with the computer.
For $x\leq y\leq z$ in $\{1,r,s\}$, any triangle with circles of sizes $x$, $y$ and $z$ which appear in an FM-triangulation of a saturated packing has edge length in the compact set
\[
[x+y,x+y+2s]\times[x+z,x+z+2s]\times [y+z,y+z+2s].
\]
Indeed, its support circle has size at most $s$ (saturation hypothesis) so that the center of a circle of size $q$ is at distance at most $q+s$ from the center of the support circle.
We can thus compute $E(T)$ and $U(T)$ using these intervals for the edge lengths of $T$.

Of course, since these intervals are quite large, we get for $E(T)$ and $U(T)$ large overlapping intervals which do not allow to conclude whether $E(T)\geq U(T)$ or not.
We use dichotomy: while the intervals are too large to conclude, we halve them and check recursively on each of the $2^3$ resulting compacts whether $E(T)\geq U(T)$ or not.
If we get $E(T)\geq U(T)$ at some step, we stop the recursion.
If we get $E(T)<U(T)$ at some step, we throw an error: the local inequality is not satisfied!

At each step, we also compute the size of the support circle\footnote{The formulas giving the size of the support circle and the distance of its center to an edge of the triangle can be rather easily obtained with computer algebra, since they follows from simple equations (equality of distances to circle centers).
They are however quite ugly and have no special interest here, so they are not given here.}.
It is an interval and if its left point is greater than or equal to $s$, we can conclude that the triangle cannot appear in a FM-triangulation of a saturated packing and we stop the recursion.

Last, we also stop the recursion if we get an $\varepsilon$-tight triangle at some step, that is, if we get a subset of the compact
\[
[x+y,x+y+\varepsilon]\times[x+z,x+z+\varepsilon]\times [y+z,y+z+\varepsilon].
\]
Indeed, the local inequality is then already ensured by Prop.~\ref{prop:local_tight}.
This point is crucial and explains why we focused on $\varepsilon$-tight triangles in Section~\ref{sec:local_tight}.
Since $E(T)=U(T)$ for tight triangles, if a compact contains the point $(x+y,x+z,y+z)$, no matter how small it is, it yields for $E(T)$ and $U(T)$ overlapping intervals which do not allow to decide whether $E(T)>U(T)$ or not: the recursion would last forever!

The whole process terminates in around $2$ minutes on our computer, examining $172091$ triangles without throwing any error.
This proves the proposition.
\end{proof}

\section{Further comments}
\label{sec:further}

Theorem~\ref{th:main} has been here proven only for specific sizes of discs.
Does the proof adapt to other cases, in particular the $9$ cases of \cite{Ken06} and the $164$ cases of \cite{FHS}?
With this issue in mind, let us here discuss the various arbitrary choices that have been made and the possible improvements that have not been necessary but could be necessary in other cases.

The first arbitrary choice is to set $U(T)=E(T)$ for the tight triangles which do not appear in the target packing.
This choice is guided by the fact that these triangles are particularly dense, but one could imagine a modification (probably slight).

The second arbitrary choice is to set the vertex potential of the apexes of the tight triangles to zero ($V_{aba}=0$ for $a\neq b$ in $\{1,r,s\}$).
There is really no other reason than simplicity.
Actually, these potentials could be treated as variables: instead of the lower bound on $m_q$ of Prop.~\ref{prop:vertex_pot}, we wouldget inequations on the vertex potential of the apexes and the $m_q$'s that shall be satisfied in order to have the global inequality~\eqref{eq:vertex}.
Further, Prop.~\ref{prop:local_tight} would be rewritten as a series of inequalities on the same variables and, in addition, $\varepsilon$.
All these inequations would yield a polytope\footnote{In $\mathbb{R}^{10}$: $6$ apex potentials, $3$ $m_q$'s and $\varepsilon$.} in which we shall choose a point, in the hope that the local inequality~\eqref{eq:local} holds on the other triangles.
This may look a bit complicated but could surely be done if necessary.

The third arbitrary choice are the constants $l_{xy}$ and $q_{xy}$ in Prop.~\ref{prop:local}.
Let us explain how we chose these values.
The rule of thumb (which could perhaps be made rigorous) is that if the local inequality works for the triangles with only one pair of circles which are not tangent, then it should work for any triangle.
We thus consider triangles $T$ with circles of size $x$ and $y$ centered on the endpoints of an edge $e$ and vary the length of $e$,  as in Fig.~\ref{fig:plot}.
We choose $l_{xy}$'s and $q_{xy}$'s so that, regardless of the size of the third circle of $T$:
\begin{enumerate}
\item $d_e(T)<0$ as soon as $e$ is longer than $l_{xy}$;
\item $U(T)$ is slightly less than $E(T)$ when $T$ is stretched.
\end{enumerate}

Last, let us describe a possible improvement, which was not necessary here but that we think it would have a role to play in other cases.
In the hope of satisfying the inequality~\eqref{eq:local}, it is in our interest to minimize as much as possible the vertex potential (Def.~\ref{def:vertex_pot}).
Consider the center $v$ of a circle of size $q$.
Around $v$, the contribution per radian of the vertex potential $U_v$ is bounded from below by the minimum over the tight triangles $\check T$ with a circle of size $q$ in $v$ of $U_v(\check T)/\widehat{v}(\check T)$.
Hence, whenever the vertex potential is larger than $2\pi$ times the absolute value of this minimum in some triangle $T$ which contains $v$, the potential of the other triangles cannot be negative enough so that the sum around $v$ becomes negative: the inequality~\eqref{eq:vertex} still holds in $v$.
We can thus {\em cap} the vertex potential, replacing $U_v(T)$ in Def.~\ref{def:vertex_pot} by $\min(Z_q,U_v(T))$, where
\[
Z_q:=2\pi \left|\min \tfrac{U_v(\check T)}{\widehat{v}(\check T)}\right|.
\]
This would only facilitate local inequality \eqref{eq:local}.
For example, in the case of this paper, we get
\begin{eqnarray*}
Z_1&=&0.0045909468722998,\\
Z_r&=&0.0037113334292734,\\
Z_s&=&0.0029789141480929.
\end{eqnarray*}
This corresponds to the dashed line in Fig.~\ref{fig:plot}.

Of course, the ultimate goal would be to have a general result holding for {\em any} saturated compact packing.
Proving particular cases should first help to better understand the role and importance of each of the many parameters involved.


\bibliographystyle{alpha}
\bibliography{53}

\end{document}